\documentclass[amstex,useAMS,epsf,amssymb,usenatbib,twocolumn]{mn2e}
\usepackage{graphicx}
\usepackage{epsfig}
\usepackage{amssymb,amsmath,color}
\usepackage{hyperref}
\usepackage{url}
\usepackage{breakurl}
\usepackage{tabularx}
\usepackage{natbib}

\newcommand{\fermilat}{{\it Fermi}-LAT}
\newcommand{\gray}{$\gamma$-ray}
\newcommand{\grays}{$\gamma$ rays}

\def\araa{ARA\&A}             
\def\apj{ApJ}                 
\def\apjl{ApJ}                
\def\apjs{ApJS}               
\def\aap{A\&A}                
\def\mnras{MNRAS}             
\def\pasp{PASP}               
\def\prd{Phys.~Rev.~D.}
\def\nat{Nature}              

\title[Search for SNe in the \textit{Fermi}-LAT data]{Search for gamma rays from SNe with
a variable-size sliding-time-window analysis of the
\textit{Fermi}-LAT data}

\author[D. A. Prokhorov, A. Moraghan, and J. Vink]{D. A. Prokhorov$^{1}$\thanks{E-mail:d.prokhorov@uva.nl},
A. Moraghan$^2$\thanks{E-mail:ajm@asiaa.sinica.edu.tw}, and J. Vink$^{1}$ \\
~\\
$^{1}$ GRAPPA, Anton Pannekoek Institute for Astronomy, University of Amsterdam, Science Park 904, 1098
XH Amsterdam, The Netherlands
\\
$^{2}$ Academia Sinica Institute of Astronomy and Astrophysics, 11F
of AS/NTU Astronomy-Mathematics Building, No.1, Sec. 4, Roosevelt Rd.,\\
Taipei 10617, Taiwan}

\date{\today}

\pagerange{\pageref{firstpage}--\pageref{lastpage}} \pubyear{2020}

\begin{document}

\maketitle

\begin{abstract}
We present a systematic search for \gray{} emission from supernovae
(SNe) in the Fermi Large Area Telescope (LAT) Pass 8 data. The
sample of targets consists of 55,880 candidates from the
\textit{Open Supernova Catalog}. We searched for \grays{} from SNe
by means of a variable-size sliding-time-window analysis. Our
results confirm the presence of transient \gray{} emission from the
sources of non-AGN classes, including transitional pulsars, solar
flares, \gray{} bursts, novae, and the Crab Nebula, which are
projected near some of these SN's positions, and also strengthen
support to the variable signal in the direction of SN iPTF14hls. The
analysis is successful in finding both short (e.g. solar flares) and
long (e.g. transitional pulsars) high flux states. Our search
reveals two new \gray{} transient signals occurred in 2019 in the
directions of optical transients that are SN candidates,
AT2019bvr and AT2018iwp, with their flux increases within 6 months
after the dates of SN's discoveries. These signals are bright and
their variability is at a higher statistical level than that of
iPTF14hls. An exploration of archival multi-wavelength observations
towards their positions is necessary to establish their association
with SNe or other classes of sources. Our analysis, in addition,
shows a bright transient \gray{} signal at low Galactic latitudes in
the direction of PSR J0205+6449. In addition, we report the results of 
an all-sky search for $\gamma$-ray transient sources. This provided 
two additional candidates to $\gamma$-ray transient sources.
\end{abstract}


\section{Introduction}

Supernovae (SNe) are luminous explosions of stars occuring during their
last evolutionary stages \citep[for a review, see][]{Bethe90,
Hillebrandt00}. The original star, called the progenitor, is either
destroyed or collapses to a neutron star or black hole. The most
recent Galactic SN observed by the unaided eye was Kepler's SN in
1604 which was brighter than stars and planets at its peak
\citep[for a review, see][]{Vink17}. The most recent extra-galactic unaided eye SN was SN 1987A
in the Large Magellanic Cloud \citep[for a review, see][]{Arnett89}.
Two mechanisms producing SNe are re-ignition of nuclear fusion in a
white dwarf star in a binary system (a Type Ia SN) or gravitational
collapse of a massive star's core (a Type II SN). The
length of time of unaided-eye visibility of Kepler's SN and SN 1987A were
several months. Given that only a tiny fraction of the stars in a
typical galaxy have the capacity to become a supernova, it is
generally accepted that supernovae occur in the Milky Way on average
about a few times every century \citep[][]{Diehl06}. The light from
the SN corresponding to the youngest known remnant SNR G1.9+0.3 in our Galaxy
would have reached Earth some time between 1890 and 1908.

Since the rate of SNe is relatively low, observations of other galaxies with
telescopes are useful to enlarge a sample of detected SNe. Optical surveys are a
powerful tool to search for SNe \citep[e.g.,][]{Law09}.
A subset of SNe which show evidence of interaction with a dense circumstellar
medium formed by a pre-SN stellar wind are also detected in radio waves
\citep[so-called radio SNe;][]{Weiler2002} and X-rays \citep[see Table 1 from][]{Dwarkadas12}.
There are theoretical models supporting that some types of SNe, such as Type IIn and
superluminous SNe surrounded by a high-density circumstellar medium,
can emit \grays{} and be detectable with modern \gray{} telescopes if
these SNe are located at distances less than 30 Mpc
\citep[][]{Murase11, Dwarkadas2013, Rachel2019}. Nearby SNe are
acknowledged to be more promising targets for searches of \gray{} emission
from SNe owing to the inverse distance-squared law of flux.
While the remnants of SNe are well established \gray{}-emitting sources
including the young, 340- and 448-year old SN remnants, Cassiopeia A and Tycho, only one
candidate to \gray{}-emitting SNe, iPTF14hls, has been proposed
since the start of the nominal science operation of the Fermi Gamma-ray Space
Telescope (FGST) in 2008 August.

The Large Area Telescope \citep[LAT;][]{Atwood09} on-board the FGST
provides unprecedented sensitivity for all-sky monitoring of \gray{}
activity. Analysis techniques applied to searches for transient
sources require different levels of detail and coverage. For
example:
\begin{itemize}
\item Searches for variable \gray{} emission from the large region of the
sky, e.g., the Galactic plane \citep[][]{Neronov12} or the entire
sky \citep[the \textit{Fermi} all-sky variability
analysis by][]{Fermi13, Fermi17}, on the time scale of months or weeks
use a measure of variability computed as, e.g., the maximum
deviation of the flux from the average value. The reduced $\chi^{2}$
of the fit of the light curve with the constant flux is another technique
which is adopted in the \textit{Fermi}-LAT catalog
\citep[][]{Fermi20} for testing about 5,000 \gray{} sources. Both of
these statistics allow tests of a large number of positions or
sources and are not computationally expensive for a single analysis.
However, these techniques have a predetermined time interval.
\item Other searches set various lengths of time intervals after a predetermined
start time (which can be the date of SN discovery) in order to search for
a \gray{} signal during one of these time intervals.
\citet[][]{Fermi15} applied such a technique to search for \gray{}
emission from 147 Type IIn SNe using three different time windows; 1
year, 6 months, and 3 months. A smaller number of sources and three
time windows allowed them to perform a dedicated likelihood analysis
for each of these sources. However, this technique is not flexible
with respect to the selection of a start time. In the paper by \citet[][]{Fermi18},
the authors applied a sliding time window technique for a search for \gray{}
emission from 75 optically detected Galactic novae in a 15 day time
window in two-day steps ranging from 20 days before to 20 days
after the optical peak, but fixing the duration of emission.
\end{itemize}

The discovery of a transient source iPTF14hls by the Intermediate Palomar
Transient Factory occurred in September 2014. iPTF14hls is
very similar spectroscopically to a Type II-P SN, but evolved slowly,
remaining luminous for over 600 days with at least five distinct
peaks in its light curve \citep[][]{Arcavi17}. The total energy
emitted in light during the first 600 days was about
$2\times10^{50}$ erg, making iPTF14hls a luminous SN. iPTF14hls is
located at a distance of 150 Mpc which exceeds the distances to
those Type II SNe from which no \gray{} emission was found.
\citet[][]{Yuan18} reported the detection of a variable \gray{}
source positionally and temporally compatible with iPTF14hls. They
found that the source starts to emit \grays{} about 300 days after
the explosion time and the emission lasts for about 850 days. The
detection of transient \gray{} emission in the direction of
iPTF14hls gives rise to a question whether \gray{} emission comes
only from unusual SNe \citep[for a review of the models for
iPTF14hls, see][]{Woosley18}. These \gray{} observational properties
require a search for similar sources accounting for both a start and duration
of emission which serve as two variables. Previous temporal analyses
of \textit{Fermi}-LAT data often have one time variable, e.g. the index of a
time interval \citep[][]{Neronov12, Fermi13, PM16}, the duration of
a time interval \citep[][]{Fermi15, Renault18}, the oscillation
period \citep[][]{FermiPG, PM17}, the Sun's position on the ecliptic
\citep[][]{Fermi14}, or the McIlwain L parameter \citep[][]{PM19}.

The search for \gray{}-emitting SNe similar to iPTF14hls is the
ultimate goal of the paper. We have developed a
variable-size sliding-time-window technique as the first step and
apply it to a search for \gray{} emission from 55,880 SNe and
related candidates from the \textit{Open Supernova Catalog}.
We stress here that this catalog contains SN candidates, as
the supernova nature of these transients is not always entirely
established. For each of these sources we assume the existence of a
time interval during which the given source is brighter than it is
before and after this time interval. By means of a likelihood
analysis, we check if the existence of such a time interval is
statistically significant and select the most significant interval
among the possible intervals for each source. If the existence of a
high flux time interval is statistically significant, then we check
if the corresponding  date of SN discovery is within the time
interval of 300 days before the \gray{} transient. By using a
variable-size sliding-time-window analysis, we found two new
candidates with flux increases within 300 days after the SN
candidate discoveries, one new variable unidentified source at a
low Galactic latitude in the direction of PSR J0205+6449, and
confirmed a number of known \gray{} transient sources, including
\gray{} bursts, solar flares, novae, and especially transitional
pulsars, revealing high flux time intervals.

\section{Observations and methods}

\fermilat{} on the FGST spacecraft is a pair-conversion telescope
which provides tracking of electrons and positrons produced by the process of pair
production occurring for \grays{} in a converter material \citep[][]{Atwood09}.
It has a large field of view ($\approx 20\%$ of the sky) and has been scanning the sky continuously
since August 2008. These two capabilities of \fermilat{} allow efficient monitoring of the \gray{} sky.
The telescope provides an angular resolution per single event of $1.0^{\circ}$ at 0.8 GeV,
narrowing to $0.5^{\circ}$ at 2 GeV, and further narrowing to $0.1^{\circ}$ above
10 GeV\footnote{\burl{http://www.slac.stanford.edu/exp/glast/groups/canda/lat\_Performance.htm}}.
At energies below $\sim$10 GeV, the accuracy of the directional reconstruction of photon events
detected by \fermilat{} is limited by multiple scattering in the tungsten converter foils
and determined by the $\sim1/E$ dependence of multiple scattering, whereas above $\sim$10 GeV,
multiple scattering is insignificant and the accuracy is limited by the ratio of silicon strip pitch
to silicon-layer spacing.
Given the angular resolution dependence with energy, we selected the optimal lower energy
limit of 0.8 GeV to tighten the point spread function (PSF) for this analysis.
We selected the upper energy limit of 500 GeV, because of the small amount of detected events
with higher energies.

We downloaded the \fermilat{} Pass 8 (P8R3) data from the Fermi
Science Support Center and used 600 weeks of the \texttt{SOURCE}
class data (evtype=128), collected between 2008-08-04 and
2020-01-30. The SOURCE event class is tuned to balance statistics
with background flux for long-duration (e.g., on the time scale of
weeks) point source analysis. We performed the data analysis using
the \texttt{FERMITOOLS} v1.2.23 package. We rejected events with
zenith angles larger than $90^{\circ}$ to reduce contamination by
albedo \grays{} from the Earth. We applied the recommended cuts on
the data quality (DATA\_QUAL$>0$ \&\& LAT\_CONFIG$==1$). We binned
the data into time intervals of one week and in three energy bands,
namely, 0.8-2.0 GeV, 2.0-5.0 GeV, and 5.0-500.0 GeV. The choice of
three energy bands instead of a single band facilitates a study of
the \gray{} sources with soft or hard photon indices, since the
signal-to-noise ratio is expected to be higher in the 1st band for
soft sources and higher in the 3rd band for hard sources. We further
binned the \fermilat{} events using the HEALPIX package into a map
of resolution $N_\mathrm{side}$ = 512 in Galactic coordinates with
`RING' pixel ordering. With these settings, the total number of
pixels is equal to 3,145,728 and the area of each pixel is
$1.3\times10^{-2}$ deg$^2$. The chosen resolution of the map is fine
enough to allow the selection of \grays{} from circular regions
around SNe. To compute the exposure, we used the standard tools
\texttt{gtltcube} and \texttt{gtexpcube2}. To correct the livetime
for the zenith angle cut, we used the `zmax' option on the command
line.

We used the \textit{Open Supernova Catalog}, an online collection of
observations and metadata for 50,000+ SNe and related candidates
\citep[][]{Guillochon17}. This catalog is freely available on the
web\footnote{\url{https://sne.space}}. The objects included in this catalog
are intended to be entirely SNe and the authors of the catalog
remove objects that have been definitively identified as other
transient types. One difference between the \textit{Open Supernova
Catalog} approach and some other catalogs is that the authors
augment the known SNe with known supernova remnants for
completeness, which are thought to be SNe but (currently) possess no known
associated transient. We extracted the positions of sources in the
sky from this catalog and computed both the total number of events within
a 0\fdg35 radius circle centered on the position of each SN and the
corresponding exposure for every week of observations. The circular
region with a 0\fdg35 radius is sufficient to accumulate a
significant number of events from the potential source, but also
relatively small to strongly suppress the contamination of signals
by \grays{} coming from numerous active galactic nuclei (AGN)
including blazars and radio galaxies, such as NGC 1275. To further
suppress the contamination, we selected SNe located at distances
larger than 1$^{\circ}$ from AGN included in the \fermilat{} catalog
\citep[][]{Fermi20}. Apart from the positions of SNe, we also
extracted discovery dates and SN types from the
\textit{Open Supernova Catalog}. We checked that the
\textit{Open Supernova Catalog} is rather uniform and the total covered
surface by our circular regions is a significant portion of the sky.

We developed a python code which performs a likelihood analysis
for finding the most statistically significant time interval of a
high flux for every selected source, and it is publicly
accessible at \burl{https://zenodo.org/record/4739389}. To search for such a
time interval, we compared two models with and without the presence
of a bright state. The ``null'' model assumes a source with a steady
flux in time. The alternative model assumes the presence of a time
interval during which a source has a flux different from that which
is before and after the bright state. Taking the exposure for each
week into account, we estimated the number of expected counts from
the source during each week and computed a Poisson probability using
the observed number of counts. The product of Poisson probabilities
for all weeks gives us a likelihood for the given model. We employed
the Test Statistic (TS) to evaluate the significance of evidence for
a bright state. The TS value is defined as
$TS=2\ln\left(\mathcal{L}(H_{1})/\mathcal{L}(H_{0})\right)$, where
$\mathcal{L}(H_{0})$ is the maximum likelihood value for the null
model and $\mathcal{L}(H_{1})$ is the maximum likelihood for the
alternative model. We considered each energy band independently from
the other two bands allowing an analysis independent on the photon
index. Since the null model represents a special case of the
alternative model, the probability distribution of the TS is
approximately a chi-square distribution with three degrees of
freedom - the difference between the numbers of free parameters of
the null and alternative models (one degree is for each energy band)
accordingly to Wilks' theorem.

\begin{figure}
\centering
\includegraphics[angle=0, width=.4\textwidth]{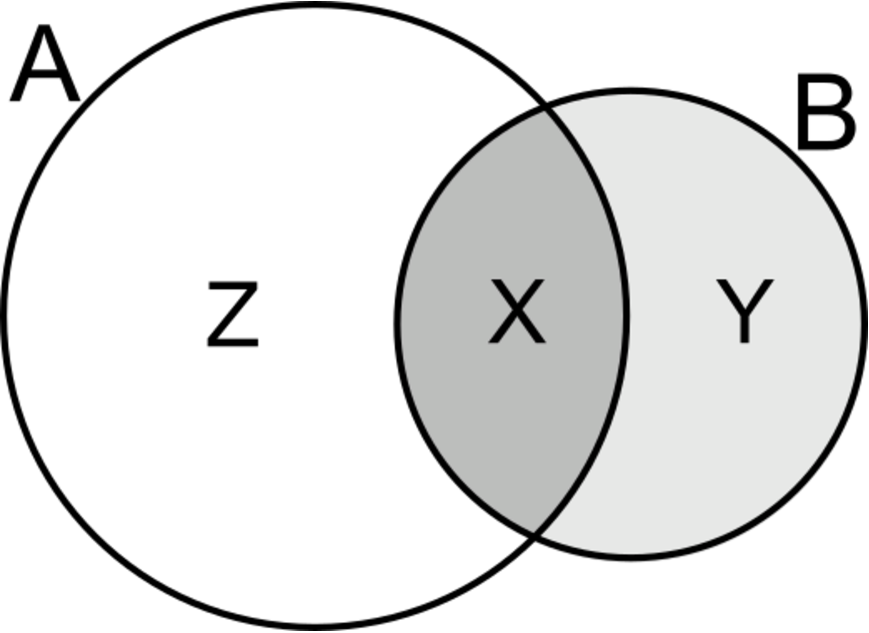}
\caption{The Venn diagram showing the relationship between the
sets, $A$ and $B$, and their subsets, $X$, $Y$ and $Z$.}
\label{vennplot}
\end{figure}

\begin{figure*}
\centering
  \begin{tabular}{@{}cc@{}}
    \includegraphics[angle=0, width=.47\textwidth]{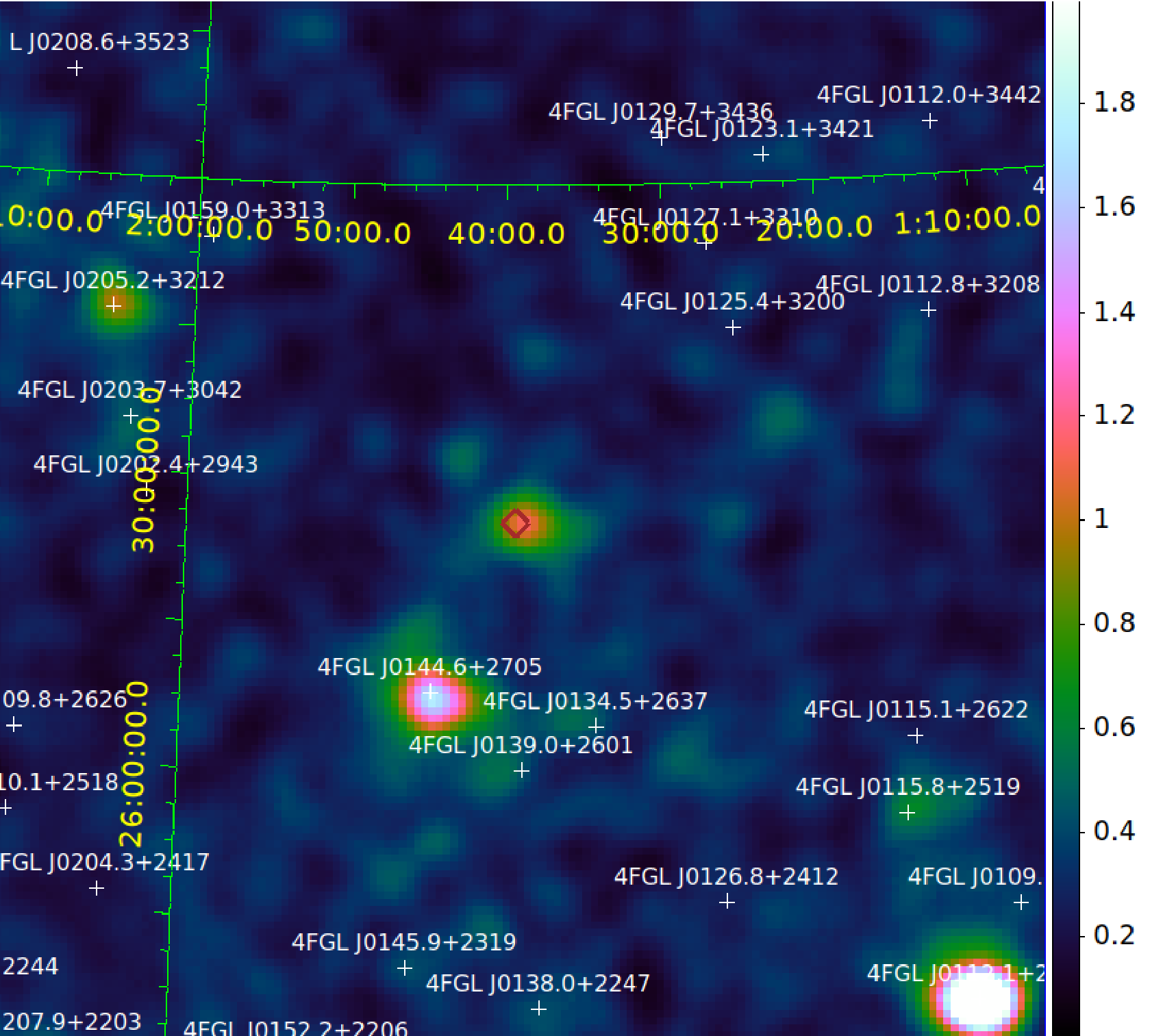}
    & \ \ \ \ \ \ \ \ \
    \includegraphics[angle=0, width=.47\textwidth]{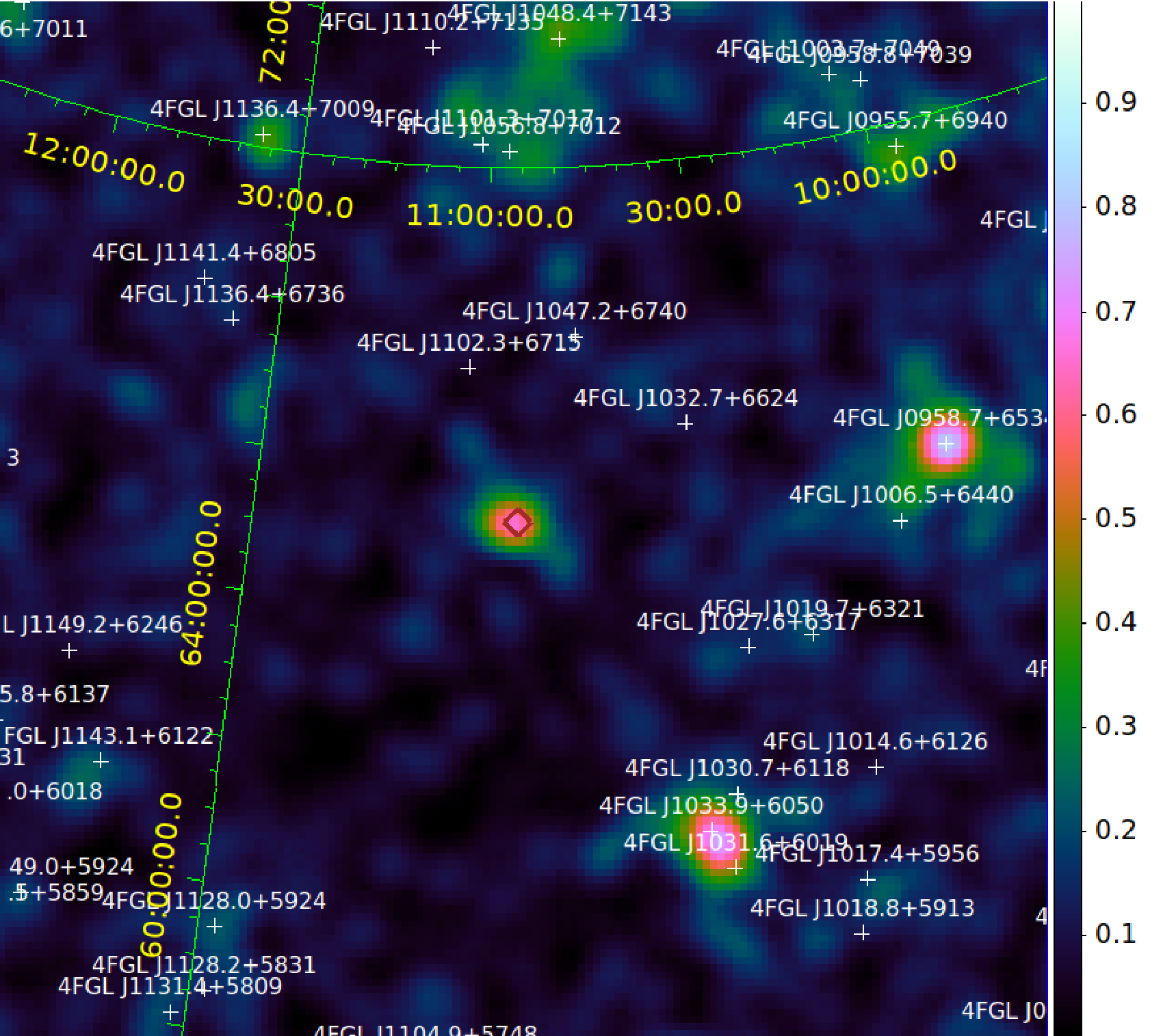} \\
   \end{tabular}
  \caption{Gaussian ($\sigma=0.5^{\circ}$) kernel smoothed count maps centered on the positions of AT2018iwp (left panel)
  and AT2019bvr (right panel). The positions of \gray{} sources from the 4FGL-DR2 catalog are shown with crosses, and
  AT2018iwp and AT2019bvr by diamonds.} \label{F1}
\end{figure*}

We searched for bright state emission in a variable-size
sliding-time window in one week steps ranging from the first week of
\fermilat{} observations to the last one of 600 weeks. Thus, the
shortest time window is one week and the longest is 599 weeks. The
time interval of a high flux state can be written as $\{i, j\}$, while the
time interval of a normal flux state can be written as $\{1, i\}\cup\{j,
600\}$, where $i$, $j$ are week numbers in the ranges of $\{1,
599\}$ and $\{i+1, 600\}$, respectively.
The variable-size sliding-time-window
analysis of 55,800 SNe took 12 days to run on the servers using 56
scripts in parallel that analyzed 1000 sources each.
We tested the algorithm on a flaring blazar and found the algorithm
correctly reveals a time interval corresponding to a given flare. We
also tested the algorithm on simulated data including a high flux
state at a given duration and found that the algorithm
successfully identifies a bright state in the simulated data. Given
that the number of weeks is 600 and thus a large number of trials,
we expected to find the time interval of a high flux for every
source at a statistical level above $3\sigma$. The performed analysis
confirmed that the average value of a statistical level for selected
sources is indeed $3.3\sigma$. We adapt a global significance level
where we indicate the significance level after taking the ``look elsewhere effect''
into account which is quantified in terms of a trial factor that is the ratio of
the probability of observing the excess in the obtained time interval
to the probability of observing it with the same local significance level
anywhere in the allowed range for a given SN position.
Below we focus on two samples of sources with the presence of
a high flux time interval:

(A) at a local significance level higher than 6.0 standard deviations,
which translates to a global significance of about
5.0$\sigma$ and

(B) at a local significance exceeding 5.0 standard deviations and
which starts within 300 days after the date of SN discovery
(however, if a SN occurred before the Fermi mission, then to be
included in this set a \gray{} signal should start during the first
weeks of the Fermi mission\footnote{It is worth to mention that it
is conceivable that the \gray{} signal from a SN occurred before the
Fermi mission can start during the mission, see
\citet[][]{Berezhko2015}}).

The criterion for classifying transient sources for the sample, A,
satisfies the convention of a $5\sigma$ global significance level.
The expected number of false detections in the sample, A, is
$55880\times1000\times(2.0\times10^{-9})\simeq0.12$, where $55880$
is the number of SNe candidates and $1000$ is the trial factor
related to the choice of time intervals. The criterion used for the
sample, B, is for checking the completeness of our sample of
transients which can be associated with SNe. The expected number
of false detections in the sample, B, is
$55880\times1000\times(5.9\times10^{-7})/(600/43)\simeq2.3$, where
$600$ is the total number of observation weeks and $43$ is the
number of weeks corresponding to $300$ days\footnote{If the
constraint on the start time of emission was not set, then the
number of spurious detections in the sample, B, would be 33.}.

The choice of a 300 day interval for the start date of \gray{}
emission is based on the fact that GeV-TeV \grays{} almost
simultaneous with the optical/infrared light curves are expected and
to be emitted in 1-10 months \citep[][]{Murase11} which motivates
searches for $\sim$0.1-1 yr transients via multi-year
\textit{Fermi}-LAT observations. In the optically thin scenario, the
\gray{} radiation time duration corresponds to the SN-shock-crossing
time through the circumstellar medium modelled as a shell of matter.
The $\gamma\gamma$ absorption on the non-thermal synchrotron photons
can lead to a delay of the onset of GeV emission by $\sim$1 month
\citep[][]{Kantzas2016}. Given that the \gray{} source appeared
about 300 days after the explosion of iPTF14hls, which was proposed
to be the first detected \gray{} emitting SN and is furthermore a
prototype of transients for our search, we generalized the condition
that emission is within 1 year to the condition that emission starts
within 300 days by covering the empirically-based iPTF14hls case.
Our search includes a search for transients within the 3-month,
6-month, and 1-year intervals \citep[e.g.,][]{Fermi15}
as subsamples. A longer delay in the onset of gamma-ray
emission is also conceivable \citep[e.g.,][]{Berezhko2015}, however
the perspective of association between a transient event and a SN is
less certain in this case and requires multi-wavelength observing
campaigns, such as those which have taken place for SN 1987A for the
last three decades. Given the lack of established GeV \gray{}
sources identified with SNe - with the possible exception of
iPTF14hls,
we followed a conservative approach assuming that the onset of GeV emission from
SNe can be at any time within the 300 day window and that the time
duration of GeV emission can be arbitrary.

We consider three sets, $X=A\cap B$ (i.e. $X$ contains only those
elements which belong to both $A$ and $B$), $Y=B\setminus X$
(i.e. $Y$ contains only elements of $B$ which are not in $X$), and
$Z=A\setminus X$ (i.e. $Z$ contains only elements of $A$ which are
not in $X$). The Venn diagram shown in Figure \ref{vennplot}
illustrates the relationship between these sets. Signals from set,
$X$, should be associated with SNe with a higher probability.
The probability of detecting one signal in the set, X, by mere
chance is
$55880\times1000\times(2.0\times10^{-9})/(600/43)\simeq8\times10^{-3}$.
Signals from the set, $Y$, can also be associated with SNe, but
their variability is at a lower statistical level. Signals from the
set, $Z$, are likely associated with other transients which are not
related to SNe. Given the search for a high (not low) flux time
interval, we considered the cases in which the data in at least one
of the three independent energy bands show the time interval with a
positive flux variation.

While performing the analysis, we found that some high flux time intervals
with a global significance above $5.0\sigma$ are associated
with strong \gray{} flares of known AGN from the \fermilat{} catalog and
located at distances (a little) larger than 1$^{\circ}$ from \textit{Open Supernova Catalog}
sources. The list of these AGN include 4C 01.02, PKS 0346-27, SBS 0846+513, Ton 599,
4C +21.35, 3C 279, PKS B1424-418, PKS 1502+106, PKS 1510-089, CTA 102, PKS 2227-08, PKS 2247-131,
and 3C 454.3. We checked that the high flux
time intervals obtained from our variable-size sliding-time-window analysis
correspond to the flares of these AGN giving us confidence in the reliability of the method.
Below we do not consider these sources since the flaring activity of these AGN have already
been reported in \textit{Fermi} Astronomer's
Telegrams\footnote{\burl{https://heasarc.gsfc.nasa.gov/docs/heasarc/biblio/pubs/fermi\_atel.html}}
including ATel \# 2328, 2584, 3452, 8319, 10931, 11251, 11542, and 11141, and these AGN are
included in the catalog \citep[][]{Fermi17} with the exception of PKS 2247-131 whose flare
was in 2018 after that publication.

\section{Results}

\begin{table*}
\centering \caption{The list of transient \gray{} signals obtained
from a variable-size sliding-time-window analysis. The second column
shows the set to which a signal corresponds. The third and fourth
columns show the Right Ascension and the Declination of a SN. The
fifth and sixth columns show the name and the discovery date of a
SN. The seventh and eighth columns show the start date and the length
of a high-\gray{}-flux state. The ninth column shows the local
significance at which the high flux state is present. The tenth
column shows whether the source is firmly identified
({\large{$\blacktriangle$}})  or possibly associated
({\large{$\vartriangle$}}) with a transient \gray{} signal. The
number in brackets (if shown) indicates how many sources from the
\textit{Open Supernova Catalog} are affected by the presence of a
given \gray{} source.}
\begin{tabular}{ | c | c | c | c | c | c | c | c | c | c| }
\hline
\# & Set & R.A. & Dec. & Name & Disc. Date & \gray{} bright  & Length  & Local. & Id./Assn. (\large{$\blacktriangle$/$\vartriangle$}) \\
 & &  (hh:mm:ss)       &  (hh:mm:ss)          &  &  yr/m/d          & state from &  (week) & signif.           &          \\
& &  &  &  &  & (yr/m/d) &  & &  \\
\hline
N01 & $X$ & 01:39:24 & +29:24:06 & AT2018iwp & 2018/11/07 & 2019/04/22 & 15 & 6.0$\sigma$ & \textbf{new} \\ 
N02 & $X$ & 10:55:33 & +65:09:55 & AT2019bvr & 2019/02/20 & 2019/05/13 & 5 & 7.8$\sigma$ & \textbf{new} \\ 
N03 & $X$ & 10:22:23 & +01:12:06 & LSQ13afs & 2013/04/29 & 2013/07/01 & 334 & 6.7$\sigma$ & PSR J1023+0038 \large{$\vartriangle$} (2) \\ 
N04 & $X$ & 11:32:32 & +27:41:56 & SN2013cq & 2013/04/27 & 2013/04/22 & 1 & $>8.0\sigma$ & GRB 130427A \large{$\blacktriangle$} (4) \\ 
\hline
N05 & $Y$ & 01:05:02 & -21:56:12 & SN2018gxi & 2018/09/28 & 2019/01/28 & 52 & 5.4$\sigma$ & \textbf{new} \\ 
N06 & $Y$ & 04:26:19 & -10:27:45 & SN2017htp & 2017/11/01 & 2017/10/09 & 1 &  5.5$\sigma$ & GRB 171010A \large{$\blacktriangle$} \\ 
N07 & $Y$ & 09:20:34 & +50:41:47 & CSS141118 & 2014/11/18 & 2015/06/01 & 109 & 5.3$\sigma$ & iPTF14hls \large{$\vartriangle$}\\ 
\hline
N08 & $Z$ & 02:10:33 & +64:07:48 & GRB 080727C & 2008/07/27 & 2017/03/13 & 4 & 6.9$\sigma$ & \textbf{new} \\ 
N09 & $Z$ & 08:38:10 & +24:53:26 & SN2018ggc & 2018/09/08 & 2018/01/08 & 1 & 7.0$\sigma$ & \textbf{new}\\ 
N10 & $Z$ & 00:42:44 & -01:29:40 & PS19iho & 2019/12/24 & 2013/12/30 & 1 & 7.6$\sigma$ & GRB 131231A \large{$\blacktriangle$} (1) \\ 
N11 & $Z$ & 05:34:31 & +22:01:00 & SN1054A & 1054/07/04 & 2011/04/11 & 1 & $>8.0\sigma$ & Crab Nebula flare \large{$\blacktriangle$} \\ 
N12 & $Z$ & 05:42:47 & +82:27:26 & AT2013kg & 2013/10/23 & 2019/06/24 & 2 & 6.7$\sigma$ & FSRQ S5 0532+82 \large{$\blacktriangle$} \\ 
N13 & $Z$ & 08:58:53 & +17:08:37 & PS19ivt & 2019/12/27 & 2009/07/27 & 1 & 6.0$\sigma$ & quiescent Sun \large{$\blacktriangle$} \\ 
N14 & $Z$ & 10:35:40 & -59:42:00 & G286.5-01.2 & -- & 2018/04/02 & 4 & $>8.0\sigma$ & Nova ASASSN-18fv \large{$\blacktriangle$} \\ 
N15 & $Z$ & 10:44:21 & +08:20:11 & MLS110526 & 2011/05/26 & 2014/08/25 & 1 & 7.0$\sigma$ &  solar flare \large{$\blacktriangle$} \\ 
N16 & $Z$ & 11:12:29 & +05:03:04 & AT2019blr & 2019/01/20 & 2017/09/04 & 1 & $>8.0\sigma$ & solar flare \large{$\blacktriangle$} (7) \\ 
N17 & $Z$ & 12:26:48 & -48:46:04 & Gaia16cdq & 2016/12/02 & 2009/02/02 & 199 & 6.3$\sigma$ & PSR J1227-4853\large{$\blacktriangle$} \\    
N18 & $Z$ & 13:19:46 & -08:25:37 & AT2018aee & 2018/03/06 & 2019/10/14 & 1 & 6.5$\sigma$ & quiescent Sun \large{$\blacktriangle$} \\ 
N19 & $Z$ & 18:26:05 & -13:03:20 & G18.45-0.42 & -- & 2012/06/04 & 51 & 6.2$\sigma$ & PSR J1826-1256 \large{$\vartriangle$} \\ 
N20 & $Z$ & 20:20:50 & +40:26:00 & DR4 & -- & 2009/07/20 & 117 & $>8.0\sigma$ & PSR J2021+4026 \large{$\blacktriangle$} \\   
N21 & $Z$ & 20:33:46 & +07:00:44 & AT2019isx & 2019/05/29 & 2016/06/20 & 1 & $>8.0\sigma$ & GRB 160625945 \large{$\blacktriangle$} \\ 
N22 & $Z$ & 21:02:03 & +42:13:55 & Gaia19eym & 2019/11/05 & 2016/06/20 & 1 & 6.0$\sigma$ & GRB 160623209 \large{$\blacktriangle$} \\ 
N23 & $Z$ & 22:14:01 & -26:55:44 & SN2010bv & 2010/04/19 & 2009/05/04 & 1 & $>8.0\sigma$ & GRB 090510A \large{$\blacktriangle$} (3) \\ 
N24 & $Z$ & 23:14:03 & -04:44:20 & PS1-14afk & 2014/12/18 & 2012/03/05 & 1 & $>8.0\sigma$ & solar flare \large{$\blacktriangle$} (2) \\ 
N25 & $Z$ & 23:35:19 & -66:11:05 & OGLE16cpa & 2016/07/12 & 2009/09/21 & 1 & $>8.0\sigma$ & GRB 090926A \large{$\blacktriangle$} (2) \\ 
\hline
\end{tabular}
\label{Tab}
\end{table*}

We present the results in Table \ref{Tab} which contains the list of sources from the three sets,
$X=A\cap B$, $Y=B\setminus X$, and $Z=A\setminus X$. We detected two new sources belonging to
the set, $X$, which can potentially be associated with SNe given that the variability of sources from
this set is at a high statistical level and that these \gray{} signals started within a 300-day time
interval after the date of a SN discovery. We detected one new transient source in the set, $Y$,
but located at an offset from the SN's position.
We also detected two new transient sources in the set, $Z$, including one source at a low Galactic
latitude and the other source likely associated with a blazar.

\subsection{Sources of the set $X=A\cap B$}
\label{sectX}

The set, $X$, contains four sources including two newly detected
ones, N01 and N02, which are possibly associated with AT2018iwp and
AT2019bvr. Given that the probability of detecting two new
sources in the set, $X$, by mere chance is
$(55880\times1000.0\times(2.0\times10^{-9})/(600/43))^2\simeq6.3\times10^{-5}$,
this constitutes $4.0\sigma$ evidence for $\gamma$-ray emission from
transient sources occurring in the directions of SN candidates. The
other two sources, N03 and N04, are associated with the already
known transient \gray{} sources, PSR J1023+0038 (N03) and GRB
130427A (N04).

The signals, N01 and N02, have not yet been associated with any
known \gray{} sources. Our analysis reveals that both these
transient signals occurred in 2019 and lasted for several weeks. The
signal, N01, started about 5 months after AT2018iwp, while the
signal, N02, started about 3 months after AT2019bvr. Apart from the
signals, N04 (GRB 130427A) and N06 (GRB 171010A; see Sect. 3.2), 5
of all the 23 signals in Table \ref{Tab} occurred within 300 days
after the SN events from the \textit{Open Supernova Catalog}. Given
that iPTF14hls is one of these 5 signals (see Sect. 3.2) and that 16
of the remaining 18 signals are firmly identified, we examined the newly
detected signals, N01 and N02, in more detail.

We performed binned likelihood analyses of the sources located at
the positions of AT2018iwp and AT2019bvr using the standard \texttt{FERMITOOLS}
package. We selected events with energies in the range from 300 MeV
to 500 GeV and with reconstructed directions within a
$15^{\circ}\times 15^{\circ}$ region of interest around each of
these two sources. We chose the photon events recorded during the
time intervals shown in Table \ref{Tab}. Figure \ref{F1} shows
Gaussian ($\sigma=0.5^{\circ}$) kernel smoothed count maps centered
on the positions of AT2018iwp and AT2019bvr and illustrates the
presence of \gray{} excesses during the corresponding time
intervals. We binned the data in 25 equal logarithmically spaced
energy intervals and used a $0\fdg1\times0\fdg1$ pixel size. To
model the Galactic and isotropic background diffuse emission, we
used the templates, \texttt{gll\_iem\_v07} and
\texttt{iso\_P8R3\_SOURCE\_V2\_v1.txt}. The other cuts applied to
the Fermi-LAT data are identical to those used in Sect. 2. We built
a complete spatial and spectral source model using point sources
from the LAT 10-year Source Catalog \citep[4FGL-DR2;][]{Ballet2020}.
Using the \texttt{gtlike} routine, we found that the \gray{} source
at the position of AT2018iwp is at an $11.3\sigma$ statistical level
and that the \gray{} source at the position of AT2019bvr is at a
$10.3\sigma$ statistical level. (We clarify that these significances
correspond to the detection of a \gray{} source at the given position
during the high flux time interval, while the significances shown in
Table \ref{Tab} are for the existence of a high flux time interval.)
It demonstrates that these signals are at a high statistical level
and also a potential for revealing new \gray{}-emitting sources
using a variable-size sliding-time-window, see also the signals,
N09 and N12, in Sects. 3.2 and 3.3.

\begin{figure*}
\centering
  \begin{tabular}{@{}cc@{}}
    \includegraphics[angle=0, width=.5\textwidth]{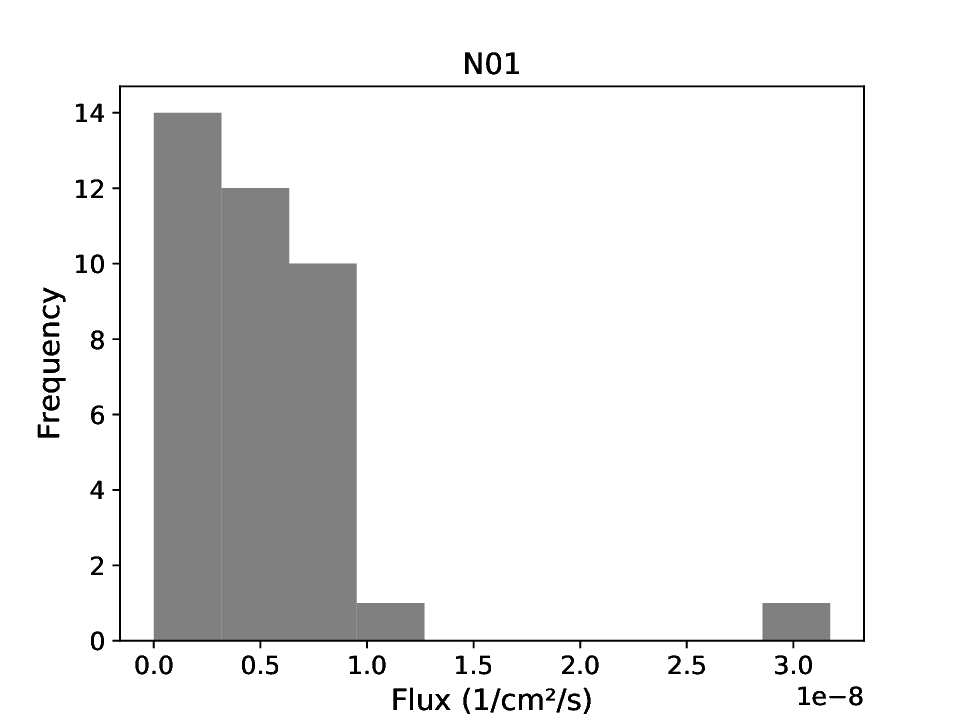}
    & \ \ \ \ \
    \includegraphics[angle=0, width=.5\textwidth]{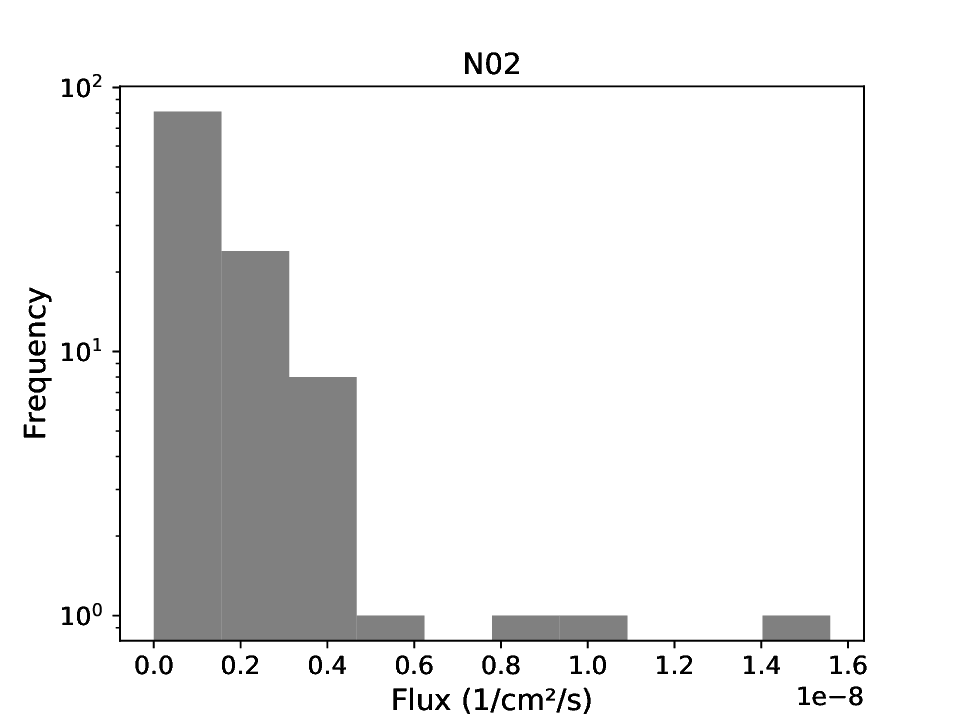} \\
  \end{tabular}
  \begin{tabular}{@{}ccc@{}}
    \includegraphics[angle=0, width=.33\textwidth]{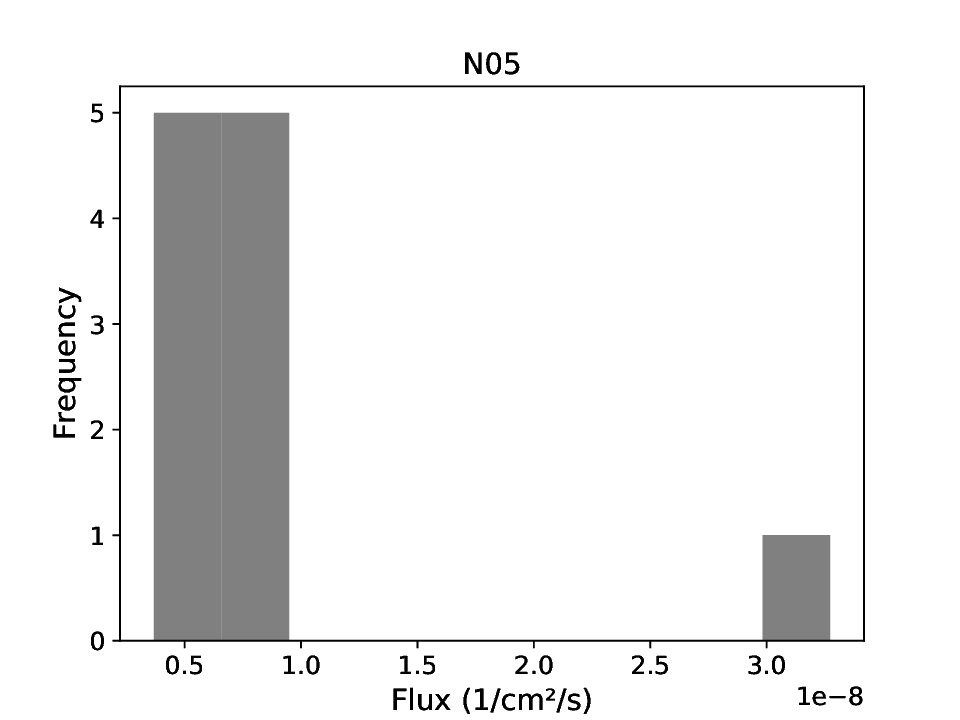}
    &
    \includegraphics[angle=0, width=.33\textwidth]{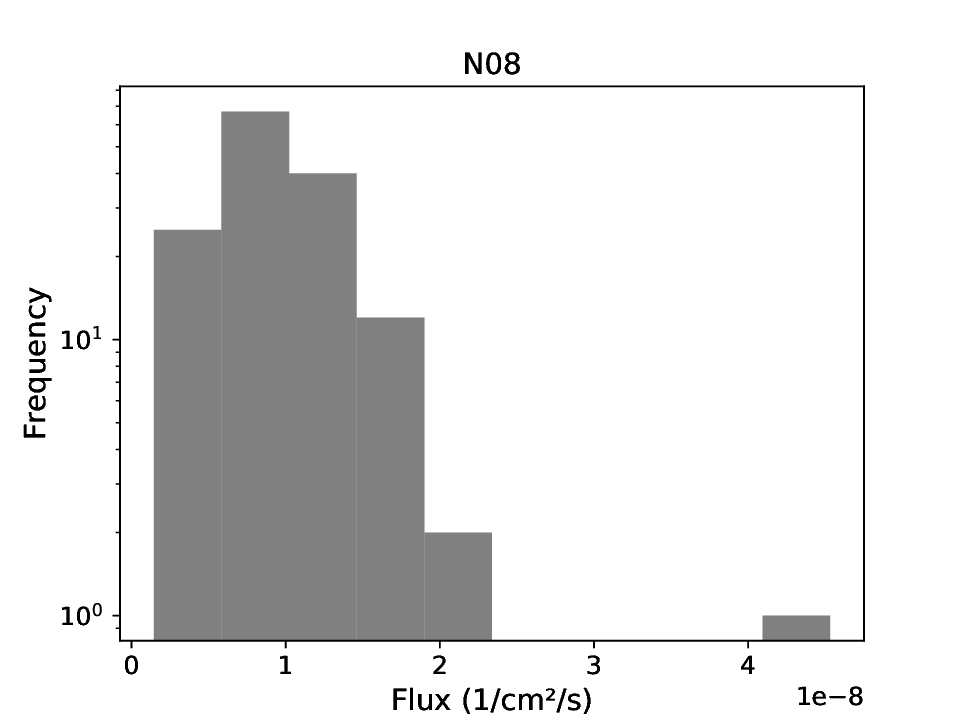}
    &
    \includegraphics[angle=0, width=.33\textwidth]{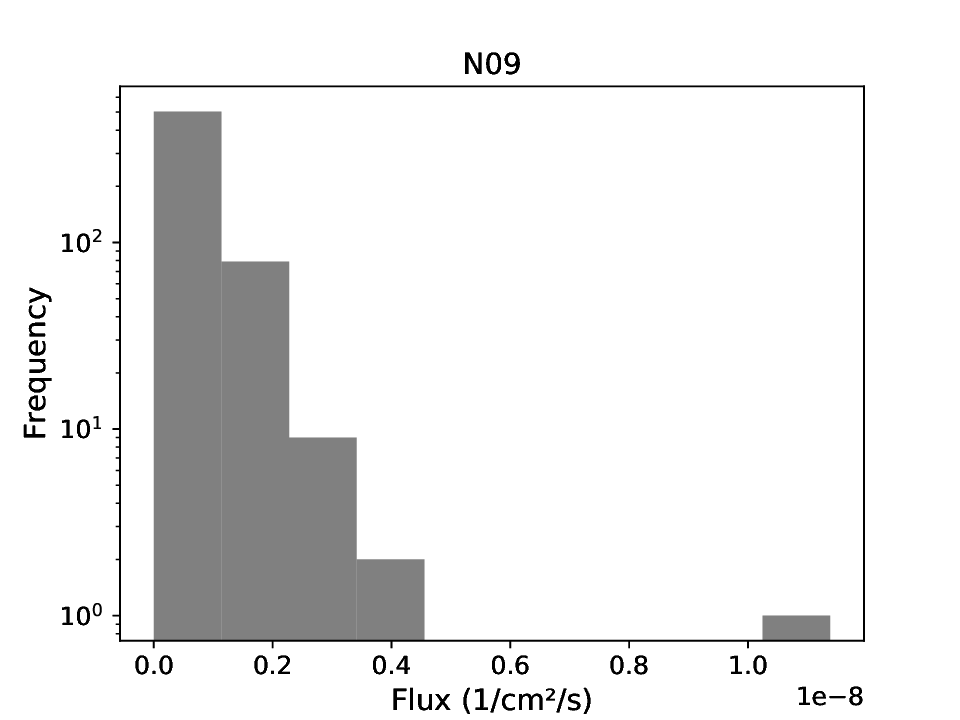}\\
  \end{tabular}
  \caption{Histograms showing the presence of bright states in the directions of AT2018iwp (N01), AT2019bvr (N02), SN2018gxi (N05),
PSR J0205+6449 (N08) and CGRBS J0837+2454 (N09).} \label{histo}
\end{figure*}

To give us confidence in the reliability of computed time windows, we examined the temporal behavior of the signals, N03 and N04,
and found that the week during which the change in flux occurred for the signal, N03, and its duration are
compatible with those which are reported by \citet[][]{Stappers2014} and \citet[][]{Xing18}
and that the week during which the
signal, N04, occurred is the same as that which is reported by \citet[][]{FermiGRB130427A}. The signal, N03, is associated
with a transitional millisecond pulsar binary, PSR J1023+0038, which changed its state from the disk-free state
to the active state of having an accretion disk in June 2013 \citep[][]{Stappers2014}.
The position of PSR J1023+0038 is at a distance of $0\fdg65$ from LSQ13afs from the \textit{Open Supernova Catalog}
and some of its \grays{} are therefore inside the aperture of $0\fdg35$ radius around the position of LSQ13afs.
Given both the multi-wavelength confirmation and the physical phenomenon, the detection of a signal, N03,
through our variable-size sliding-time-window analysis is not surprising, but the signal is associated with PSR J1023+0038.
The \gray{} burst, GRB 130427A, identified with the signal, N04, had the high fluence, highest-energy photon (95 GeV),
long \gray{} duration (20 hours), see \citet{FermiGRB130427A}. GRB 130427A is listed as a SN in the the \textit{Open Supernova Catalog}
and is indeed associated with a Type Ic SN, SN 2013cq \citep[][]{Xu2013, Melandri2014}.

Since the performed analysis establishes the most significant
bright state in the flux evolution with time, the presence of a
number of bright-flux states for a given position in the sky is not
excluded. To check if the new \gray{} transient signals shown
in Table \ref{Tab} come from the sources producing multiple
flares, we constructed a histogram for each of these five positions.
For this purpose, we binned both the counts and exposures in the
time intervals with length taken from Table \ref{Tab} in such
a way that one of these bins contains the bright state, and used the
sum of fluxes over the three energy bands. Figure \ref{histo}
shows the computed histograms. We found that the bright state
for each of the five positions corresponds to the highest flux in
the histogram. Only the histogram for N02 indicates a possible
extended ``tail'' to high fluxes containing a few bright events,
while the other four signals show a single bright signal.
The two events with high fluxes of $8.1\times10^{-9}$ and $9.4\times10^{-9}$
ph cm$^{-2}$s$^{-1}$ for N02 occurred 15 weeks after and 3.7 years before,
respectively, the brightest event which lasted for 5 weeks.
To study the possible existence of a secondary bright state in N02, we
removed the most significant bright state from the data and after
that re-ran a sliding-time-window analysis. This additional analysis
did not result in any identification of a new bright state at a
local significance level above 5 standard deviations.

\subsection{Source of the set $Y=B\setminus(A\cap B)$}

The set, $Y$, contains three transient signals including the signals identified with GRB 171010A and iPTF14hls.
GRB 171010A is in the catalog of GRBs detected by \fermilat{} \citep[][]{GRBcat2019} and SN 2017htp, a Type Ib/c
core-collapse SN, is associated with the long GRB 171010A \citep[][]{Melandri2019}.
The connection between long-duration GRBs, such GRBs 130427A and 171010A, with Type Ic core-collapse SNe
is well established \citep[][]{Woosley2006} and the presence of these two SNe in Table \ref{Tab} is physically
motivated. Given that two signals from the sets, $X$ and $Y$, are associated with long-duration GRBs, 130427A and 171010A,
and that the transient signal from PSR J1023+0038 is not associated with (and is even at a significant spatial
offset from) a corresponding SN, only 5 transient \gray{} signals in Table \ref{Tab} remain to be explained.

The signal, N05, has not yet been associated with any known \gray{}
sources. We performed binned likelihood analysis of the source
located at the position of a corresponding SN, SN2018gxi. The
details of this likelihood analysis are similar to those described
in Sect. 3.1. The analysis shows the presence of a \gray{} source at
a statistical significance of $6.1\sigma$ during the selected time
interval. However, we also found that the best-fit position of this
\gray{} source is at (RA, Dec)=(16.14$^{\circ}$, -22.22$^{\circ}$)
which is at an offset of $0\fdg3$ and the significance of a source
at this position is $8.8\sigma$. The difference in log-likelihood
values is 20 and is thus significantly exceeds the value of 11.6/2
corresponding to 0.3\% (that is $\chi^{2}$ at 2 degrees of freedom divided by 2).
Given this evidence for the presence of a spatial offset from the
position of SN 2018gxi and the fact that the variability of signals
from  the set, $Y$, is less significant than that of signals from
the set, $X$, we consider the probability that the signals, N01
and/or N02, are associated with transient events from the
\textit{Open Supernova Catalog} is higher. However, while the
transient event, SN 2018gxi, is associated with a Type II SN in the
\textit{Open Supernova Catalog}, the transient events, AT2018iwp and
AT2019bvr, are indicated as candidates to
SNe\footnote{AT2018iwp is classified as a transient
associated with AGN activity in
\burl{https://lasair.roe.ac.uk/object/ZTF18acakour/}}. By testing
55,880 positions, we found only three sources, AT2018iwp, AT2019bvr,
and SN 2018gxi, in addition to iPTF14hls which brightened within 300
days after their discovery date.

\subsection{Sources of the set $Z=A\setminus(A\cap B)$}

The set, $Z$, is the largest of the three sets and contains 18
\gray{} signals. Two of these signals are newly detected, while the
remaining signals are firmly identified.

The newly detected \gray{} signals, N08 and N09, are in the
directions of GRB 080727C and SN2018ggc and started in March 2017
and January 2018, respectively. Given that the large time gap between the
discovery date and the start time of \gray{} signals, their
associations with SNe are very unlikely. We therefore searched for blazars
located at nearby positions in the sky among the sources in the
Candidate Gamma-Ray Blazar Survey (CGRaBS) source catalog
\citep[][]{Healey2008}. We found that CGRaBS J0837+2454 is in the
proximity of SN 2018ggc (N09) in the sky. We performed a binned
likelihood analysis (with a similar setup to those described in
Sect. 3.1) to search for a \gray{} source and found that a new
\gray{} source at the position of SN 2018ggc is at a $7.9\sigma$ significance level (or
$8.1\sigma$ if the position of CGRaBS J0837+2454 is adopted). The
difference in log-likelihood values for the analyses which adopt different
positions of a new source is 1.8 and the position of CGRaBS
J0837+2454 is within a $2\sigma$ contour from the best-fit position,
$\Delta\ln\mathcal{L}=1.4$ (that is $<5.99/2$).
We therefore associate the signal, N09, with CGRaBS J0837+2454.

\begin{figure}
\centering
\includegraphics[angle=0, width=.47\textwidth]{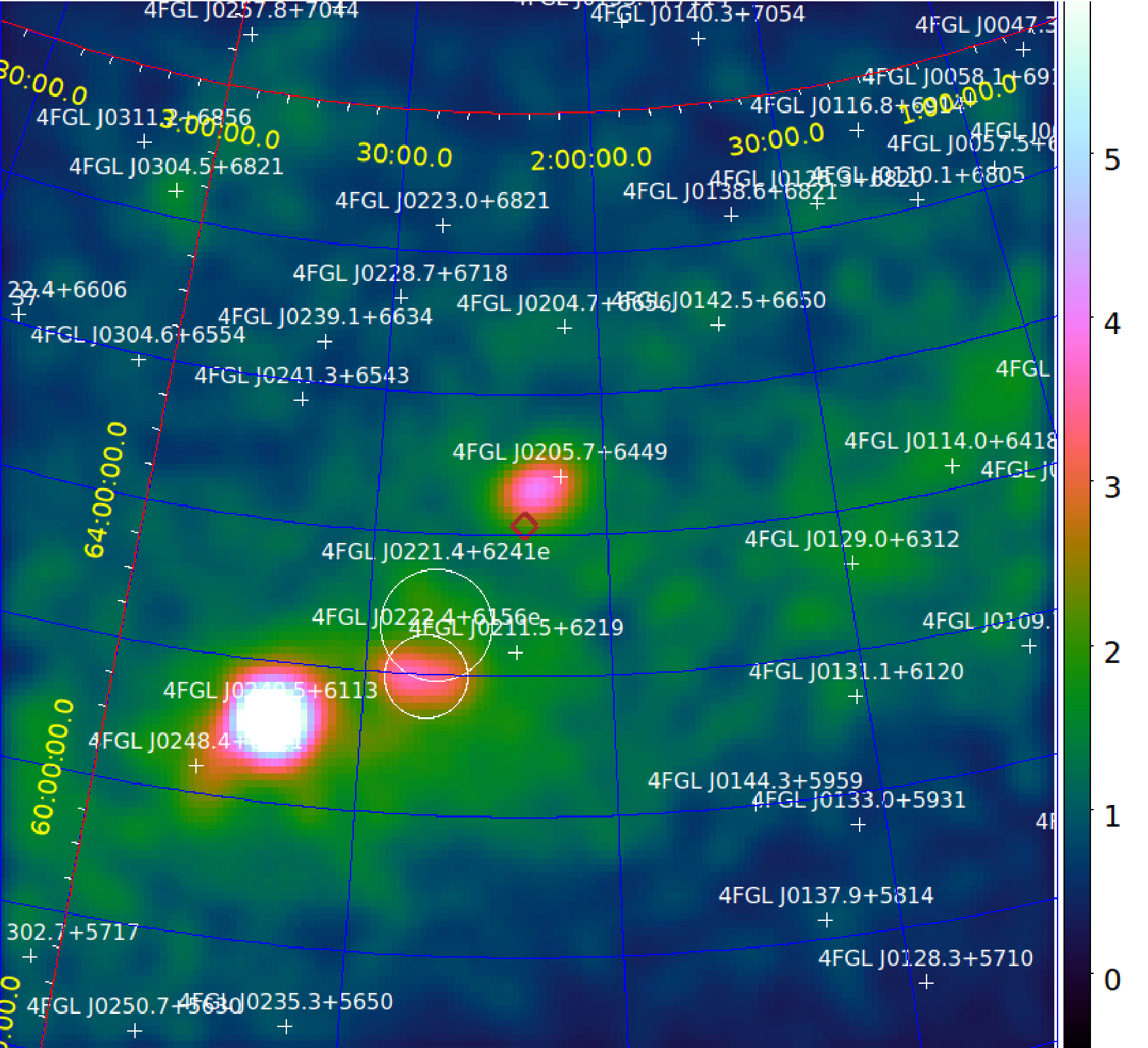}
\caption{Gaussian ($\sigma=0.5^{\circ}$) kernel smoothed count map
centered on the position of GRB 080727C shown by a diamond. The
positions of \gray{} sources from the 4FGL-DR2 catalog are shown
with crosses.} \label{F2}
\end{figure}

To study a signal, N08, in detail, we also performed a binned
likelihood analysis. In contrast with N01 and N02, N08 is located at
a low Galactic latitude of $3^{\circ}$. Figure \ref{F2} shows a
Gaussian ($\sigma=0.5^{\circ}$) kernel smoothed count map
corresponding to N08 and illustrates the presence of a \gray{}
excess during the high flux time interval. The position of a \gray{}
excess is shifted from the center of the count map towards the
position of 4FGL J0205.7+6449 which is identified with PSR
J0205+6449. Located at the center of supernova remnant/pulsar wind
nebula 3C 58 at a distance of about 3.2 kpc, PSR J0205+6449 is a
65-millisecond young rotation-powered pulsar. We found that the
\gray{} source at a position of 4FGL J0205.7+6449 is detected at a
$13.7\sigma$ statistical level during the high \gray{} flux
interval, with a \gray{} flux corresponding to $3.7\pm0.4$ times the
flux level from the 4FGL-DR2 catalog. Given that the number of
sources belonging to the Galactic plane, $\lvert b
\rvert<10^{\circ}$, in Table \ref{Tab} is only 5 which include the
Crab Nebula, Nova ASASSN-18fv, PSR J1826-1256, PSR J2021+4026, and
the \gray{} source responsible for a high flux state, N08, the
transient \gray{} signal, N08, is therefore a particularly
interesting source for further investigation and identification.

The set of \gray{} signals, $Z$, also includes well established \gray{} transient phenomena including

(i) the solar flares that occurred on 2012 March 7, 2014 September 1, and 2017 September 10 \citep[see,][for more details]{flare2012, flare2014, flare2017};

(ii) \gray{} bursts including GRB 090510A, GRB 090926A, GRB 131231A, GRB 160625945, and GRB 160623209 \citep[see,][]{GRBcat2019};

(iii) the Crab Nebula \gray{} `superflare' that occurred on April 12, 2011 \citep[][]{crabflare};

(iv) Nova ASASSN-18fv (ATel \#11546 on 2018 April 18)\footnote{Apart from this nova, we also identified Nova Mon 2012 (Atel \#4310)
in the aperture around AT2020qg among the bright state signals revealed above a global 4 sigma significance level};

(v) the quiescent Sun that passes through the corresponding regions in 2009 and 2019 during solar
minima when the quiescent \gray{} emission of the solar disk is strong \citep[see, e.g.,][]{sun2011, Fermi14};

(vi) the transitional redback PSR J1227-4853 with a transition time of 2012 November 30 \citep[][]{Johnson2015},
the variable PSR J2021+4026, which decreases in flux on 2011 October 16 \citep[][]{psrj2021}, and also
PSR J1826-1256 which has already been discussed by \citet[][]{Neronov12} and \citet[][]{Fermi17}.

(vii) FSRQ S5 0532+82 which is a \gray{}-emitting blazar announced in ATel \#12902 in July 2019.

Given that the cases (i-vii) correspond to known physical phenomena,
this fact confirms that our variable-size sliding-time-window analysis
with two temporal variables leads to sensible results and allows a search
for new transient \gray{} signals. It gives us confidence in newly
detected transient signals, particularly N01, N02, and N08.

In Table\ref{Tab}, there are only 3 highly significant
transients which have not yet been identified, including N01, N02,
and N08 (if N09 which is likely associated with a blazar CGRaBS
J0837+2454 is excluded). Given that there are 14 300-day intervals
in this data set, we find that the the chance to observe the onset
of a transient event within 300 days after the corresponding SN
explosion for two cases and not to observe it for the third case is
$3\times13/14^3\sim1.4\%$ (under the assumption that transients are
rather short which is valid for sources, N01, N02, and N08). It
means the association of these transients with SNe in time is
significant only at a $2.5\sigma$ level and thus requires further
investigation by means of an analysis of archival multi-wavelength
data for the purpose of providing us with, or questioning, their
association with SNe.

To verify the applicability of the trial factor derived in
Sect. 2 to the transient signals, N01, N02, N05, N08, and N09, we
simulated 1000 \gray{} light curves for each under the assumption of
a steady mean flux and using the given distribution of exposures
over 600 weeks and also the \texttt{numpy.random.poisson}
routine to draw counts from the Poisson distribution. We
found that the average values of local statistical levels from the
performed simulations are 3.4$\sigma$, 3.4$\sigma$, 3.2$\sigma$,
3.6$\sigma$, and 3.3$\sigma$ for N01, N02, N05, N08, and N09,
respectively, while the corresponding standard deviations are
0.4$\sigma$. These average values are compatible with those which we
derived in Sect. 2 by means of another method. We checked and found
that our classification of these transient signals into the sets, X,
Y, and Z, done in Sect. 3 is valid. By applying the sliding-time-window
analysis we found that none of these 5000 simulated steady-state
light curves results in a transient signal with a global
significance level above 5$\sigma$ which is in line with the
expectations.


\section{All-sky search for transients}

To apply our method for an all-sky search for transients, we
generated 111760 random positions uniformly distributed over the
sky. The number of random positions is selected to cover a
significant fraction of the sky, $111760\pi\times0.35^{2}=43009$
square degrees. The surface of the entire sky is
$4\pi\times(180/\pi)=41254$ square degrees. Moreover, given that
$111760/2=55880$, we can divide the set of sources at random
positions in two equal subsets, RS1 and RS2. For both these subsets
we used the same criterion as that used for the sample, A. The
sources which satisfy this selection criterion are listed in Table
\ref{Tab2}.

\begin{table*}
\centering \caption{The list of transient $\gamma$-ray signals
obtained from an all-sky variable-size sliding-time-window analysis.
This list contains signals which are additional to those reported in
Table \ref{Tab}. The second column shows the set to which a
signal corresponds. The third and fourth columns show the Right
Ascension and the Declination of a random source. The fifth and
sixth columns show the start date and the length of a
high-$\gamma$-ray-flux state. The seventh column shows the local
significance at which the high flux state is present. The eighth
column shows whether the source is firmly identified
({\large{$\blacktriangle$}})  or possibly associated
({\large{$\vartriangle$}}) with a transient $\gamma$-ray signal.}
\begin{tabular}{ | c | c | c | c | c | c | c | c| }
\hline
\# & Set & R.A. & Dec. & \gray{} bright  & Length  & Local. & Id./Assn. (\large{$\blacktriangle$/$\vartriangle$}) \\
 & &  (hh:mm:ss)       &  (hh:mm:ss)     & state from &  (week) & signif.           &          \\
 &  &                 &                 & (yr/m/d) &  & &  \\
\hline
R01 & RS2 & 01:04:53 & -22:22:19  & 2019/01/21 & 53 & 6.9$\sigma$ & \textbf{new}, N05 (Table \ref{Tab}) \\ 
R02 & RS1 & 02:43:51 & +61:04:31  & 2014/02/17 & 176 & 7.4$\sigma$ & LS I +61$^{\circ}$303 $\blacktriangle$ \\ 
R03 & RS1 & 04:42:09 & +47:15:01  & 2018/04/30 & 1 & 8.0$\sigma$ & Nova V392 Per $\blacktriangle$ \\ 
R04 & RS1 & 06:38:47 & +06:03:23  & 2012/06/18 & 2 & 6.2$\sigma$ & Nova V959 Mon 2012 $\blacktriangle$ \\ 
R05 & RS2 & 07:59:46 & -56:39:30  & 2008/09/15 & 1 & $>8.0\sigma$ & GRB 080916C $\blacktriangle$ \\ 
R06 & RS1 & 20:22:47 & +20:38:10  & 2013/08/12 & 2 & $>8.0\sigma$ & Nova V339 Del 2013 $\blacktriangle$ \\ 
R07 & RS1 & 13:54:18 & -59:07:33  & 2013/11/18 & 10 & 6.0$\sigma$ & Nova V1369 Cen 2013 $\blacktriangle$ \\ 
R08 & RS2 & 17:54:12 & +24:59:53  & 2014/02/03 & 21 & 6.9$\sigma$ & 4FGL J1753.9+2443 $\vartriangle$ \\ 
R09 & RS1 & 18:20:50 & -28:01:38  & 2016/11/07 & 2 & $>8.0\sigma$ & Nova V5856 Sgr $\blacktriangle$ \\ 
R10 & RS1 & 19:35:44 & -55:01:35  & 2016/10/10 & 18 & 6.9$\sigma$ & \textbf{new}  \\ 
R11 & RS2 & 21:00:27 & -08:55:08  & 2014/02/03 & 1 & 6.1$\sigma$ & GRB 140206B $\blacktriangle$ \\ 
R12 & RS2 & 21:03:09 & +45:53:26  & 2010/03/08 & 3 & $>8.0\sigma$ & Nova V407 Cyg 2010 $\blacktriangle$ \\ 
R13 & RS1 & 21:12:09 & +37:32:07  & 2013/07/29 & 7 & 7.3$\sigma$ & 2FAV J2111+37.6 $\vartriangle$ \\ 
R14 & RS2 & 21:29:44 & -14:55:29  & 2013/02/04 & 1 & 6.3$\sigma$ & solar event, 2013 Feb 08 $\blacktriangle$ \\ 
\hline
\end{tabular}
\label{Tab2}
\end{table*}

The all-sky search allows us to confirm a number of already
reported non-AGN transient sources in addition to those reported in
Table \ref{Tab}. These sources include six novae, including
V392 Per 2018 (ATel \#11590 in May 2018), V959 Mon 2012, V339 Del
2013, V1369 Cen 2013, V5856 Sgr 2016 \citep[][]{KwanLok}, and
V407 Cyg 2010 \citep[see also][]{Fermi18}. These sources also
include two $\gamma$-ray bursts, GRB 080916C and GRB 140206B
\citep[see][]{GRBcat2019}, and a $\gamma$-ray binary, LS I
+61$^{\circ}$303\footnote{The all-sky search revealed two signals at 
lower local significances of 5.95$\sigma$ and 5.97$\sigma$. 
The former signal is from GRB 090902B \citep[][]{GRB090902B}. 
The latter signal is at (RA, Dec)=(238.695 deg, -2.761 deg) near the position 
of a SN candidate, Gaia17adr, and started in July 2017 within 300 days 
after the SN event.}. 
This $\gamma$-ray binary is known for its
long-term $\gamma$-ray variability associated with a superorbital
period and the start of a high-$\gamma$-ray-flux state on the date
indicated in Table
\ref{Tab2} corresponds to the maximum after 1667 days
(i.e., the superorbital period) since the previous maximum reported
by \citet[][]{superorb}. The signal, R14, is due to solar
emission, since it coincides both in time and position with the Sun.
The quiescent $\gamma$-ray emission from the Sun is maximal during
the solar minimum \citep[][when the heliospheric flux of Galactic
cosmic-rays is maximal]{sun2011} and it is consistent with the
two (solar track) signals in 2009 and 2019 listed in Table
\ref{Tab}. The $\gamma$-ray signal, R14, occurred in 2013, i.e.
during the solar maximum, however the authors are unaware of
any solar flare on 
2013 February 8\footnote{\burl{https://www.spaceweatherlive.com/en/archive/2013/02/08/xray.html}}. 

Apart from these known $\gamma$-ray transients, the search also suggests that the Geminga
pulsar shows a long-term variability at a local significance level
of $6.5\sigma$. Given that Geminga is very bright in $\gamma$ rays,
systematic errors, which are not taken into account in our analysis,
might exceed statistical ones and this source thus requires a
dedicated analysis and is not included in Table \ref{Tab2}
\citep[see also][]{Pshikov2013}.

The subset, RS1, contains four unidentified signals, R10, R13,
N02, and N08, while the subset, RS2, contains three unidentified
signals, R01, R08, and N08. So the entire set contains six unidentified
signals. The probability of detecting (1) four unidentified sources
in the subset, RS1, by mere chance is
$(55880\times1000.0\times(2.0\times10^{-9}))^4\simeq1.6\times10^{-4}$
and (2) three unidentified sources in the subset, RS2, by mere
chance is
$(55880\times1000.0\times(2.0\times10^{-9}))^3\simeq1.4\times10^{-3}$.
The probability of detecting six unidentified sources in the entire
set is
$(111760\times0.75\times1000.0\times(2.0\times10^{-9}))^6\simeq2.2\times10^{-5}$,
where the factor of 0.75 is introduced in order to take the sources
overlapping in the two subsets into account. This constitutes
$4.2\sigma$ evidence for the $\gamma$-ray emission from unidentified
transient sources in the entire set. Two of these six sources, R08
and R13, are present as unidentified sources, 4FGL J1753.9+2443 and
2FAV J2111+37.6, in the Fermi-LAT 4FGL and 2FAV catalogs,
respectively. The performed all-sky search revealed the signal, R01,
associated with the signal, N05, from the set, Y, but at a higher
statistical level. It is owing to the fact that the position of R01
is closer than that of N05 ($0\fdg19$ vs $0\fdg33$) to the best-fit
position of this candidate $\gamma$-ray source.

The sample, A, contains four unidentified signals and this number is compatible 
with those for the sub-sets, RS1 and RS2. It constitutes $3.8\sigma$ evidence for the 
$\gamma$-ray emission from unidentified transient sources in the sample, A. To establish the origin of these signals and to check whether some of
them, N01 and N02, are indeed due to SN events (see Sect. 3.1), simultaneous and 
follow-up multiwavelength observations of all these unidentified candidate sources 
are required in addition to the existing optical surveys.

\section{Summary}

We developed a variable-size sliding-time-window analysis and
applied it to search for transient \gray{} signals from 55,880 SNe
listed in the \textit{Open Supernova Catalog} using 11.5 years of
the \fermilat{} data. In this paper, we reported four new \gray{}
transient signals revealed by means of a variable-size
sliding-time-window technique with global confidence above $5\sigma$
each. We labeled these four transient \gray{} signals as N01, N02,
N08, and N09 in Table \ref{Tab}. The \gray{} signals, N01 and N02,
occurred in 2019 in the directions of SN candidates, AT2018iwp and
AT2019bvr, with their flux increases within 6 months after the
reported dates of the SN candidates' discoveries. Given the
probability of detecting two new sources in the set, $X$, by mere
chance (see Sect. \ref{sectX}), this constitutes
$4.0\sigma$ evidence for $\gamma$-ray emission from transient
sources occurring in the directions of SN candidates. We obtained a
strong detection of \gray{} sources at these two positions during
the high flux time intervals at $11.3\sigma$ and $10.3\sigma$
statistical levels. The \gray{} signal, N08, occurred in 2017 and
detected at $13.7\sigma$ during the high flux time interval
corresponds to a \gray{} source at a low Galactic latitude in the
direction of PSR J0205+6449. The fourth new transient signal
occurred in 2018 and is likely due to \gray{} activity of a blazar,
CGRaBS J0837+2454. The transient \gray{} signal tentatively
associated with SN iPTF14hls by \citet[][]{Yuan18} is present in
Table\ref{Tab}, but its significance provided by a variable-size
sliding-time-window analysis is lower than those of \gray{} signals,
N01 and N02.

Among the 22 signals provided by our variable-size
sliding-time-window analysis with global confidence above $5\sigma$
(see the sets, $X$ and $Z$), we found that 17 of them are owing to
the well known astrophysical phenomena observed in
\grays{}\footnote{Given the comment by \citet[][]{Fermi17}, the
source, PSR J1826-1256, requires a dedicated analysis of its
identification.}, such as GRBs, solar flares, transitional pulsars,
novae, flares from the Crab nebula, and the moving quiescent Sun.
The developed analysis proved to be reliable in finding both short
(e.g. solar flares) and long (e.g. transitional pulsars) bright
states. This fact indicates that the three new transient \gray{}
signals, N01, N02, and N08, are most likely due to astrophysical
phenomena and therefore require further investigation for their
identification. We also performed an all-sky search for
$\gamma$-ray transient sources. It resulted in two new signals, R01
and R10. The total numbers of transient $\gamma$-ray signals from
both these analyses is 37, and 8 of them require identification.
This deserves an exploration of existing archival multi-wavelength
observations.

\section{Data availability}

\textit{Fermi}-LAT data analyzed in this paper are publicly
distributed by the LAT team and can be downloaded from the LAT Data
Server. The python code developed in this paper and used to produce
the results of the paper is publicly accessible at \burl{https://zenodo.org/record/4739389}.

\section{Acknowledgements}

We are grateful to the referee for the constructive suggestions that
helped us to improve the manuscript. Computations were performed on 
the computational facilities belonging to the ALMA Regional Center Taiwan, 
Academia Sinica, Taiwan.

\end{document}